# Efficient Anchor Point Deployment for Low Latency Connectivity in MEC-Assisted C-V2X Scenarios

Pablo Fondo-Ferreiro ◉, Felipe Gil-Castiñeira ◉, Francisco Javier González-Castaño ◉, David Candal-Ventureira ◉, Jonathan Rodriguez ◉, *Senior Member, IEEE*, Antonio J. Morgado ◉, and Shahid Mumtaz ◉, *Senior Member, IEEE*

*Abstract*—Next-generation cellular networks will play a key role in the evolution of different vertical industries. Low latency will be a major requirement in many related uses cases. This requirement is specially challenging in scenarios with high mobility of end devices, such as vehicular communications. The Multi-Access Edge Computing (MEC) paradigm seeks to satisfy it. In this article we propose the dynamic deployment of anchor point network functions at edge locations and the assignment of terminals to these anchor points with the joint objective of minimizing communications latency and reducing network overhead. We formally define the problem as a multi-objective optimization and also propose a novel heuristic greedy algorithm for approximating the solution. This algorithm compares favorably with baseline and state-of-the-art strategies for latency minimization while reducing the overhead caused by network reconfigurations.

## I. INTRODUCTION

**5G** NETWORKS are expected to support not only human-centric communications but also different vertical use cases, which pose significant requirements to 5G networks in terms of throughput, latency and availability [1]. In order to satisfy such requirements in a cost-effective manner, 5G networks have been designed for flexibility and adaptability. To a great extent, flexibility is provided by the softwarization and virtualization paradigms, which enable novel approaches such as network slicing and Multi-Access Edge Computing (MEC). The requirements for 6G networks are even more challenging, considering the use of cell-less access networks, or novel usage scenarios such as ubiquitous mobile broadband, ultra-reliable low-latency broadband communication, and massive ultrareliable low-latency communications [2].

One of the verticals that may greatly benefit from next-generation cellular networks is the automotive sector [3]. Vehicular communications or Vehicle-to-Everything (V2X) have gained attention in the last few years on the path towards connected and autonomous vehicles. Some applications using vehicular communications are collision avoidance, road status monitoring, vehicle traffic optimization and infotainment [4]. Cellular networks provide global and ubiquitous connectivity to vehicles, supporting their mobility with guaranteed quality of service leveraging the centralized orchestration of the network. 3GPP standardized Cellular V2X (C-V2X) communications in release 14 [5] as LTE-V2X, evolving them to 5G New Radio (NR)-V2X [6]. In 6G networks, the role of C-V2X will be even more relevant, addressing the challenges of connected and autonomous vehicles [7].

The need for low latency in vehicular communications (e.g., safety-related applications) has been identified as a major limiting factor for using LTE networks [8]. Therefore, recent works on low-latency vehicular communications are being focused on MEC solutions for 5G networks [9], [10], [11], [12]. In this line, in [13] we proposed a dynamic optimization framework for the deployment of anchor points in MEC locations with the sole objective of seamlessly minimizing the latency perceived by the end users. Anchor points are the main data plane network functions in cellular network architectures (i.e., S/P-GWs in 4G and UPFs in 5G). The role of anchor points in the MEC architecture is to support user plane traffic steering to the intended MEC applications in the data network [14]. The framework relies on our SDN solution for transparent Session and Service Continuity (SSC) in dynamic MEC [15], which allows relocating the serving MEC for each terminal without disrupting its communications. Note that the deployment of anchor points at MEC locations themselves is one of the most common options for MEC architectures, as reported by ETSI both for 4G [16] and 5G [14] scenarios. The connection between the MEC host and the closest base stations will typically take place through high-bandwidth low-latency wired links (e.g., optical fiber).

In this work we extend that dynamic optimization framework with the objective of not only minimizing the latency perceived

Manuscript received 20 March 2023; revised 6 June 2023; accepted 15 July 2023. Date of publication 19 July 2023; date of current version 19 December 2023. This work was supported in part by Xunta de Galicia (Spain) under Grants ED481B-2022-019, ED431C 2022/04, and IN854A 2020/01 (Factory competitiveness and electromobility through innovation, FACENDO 4.0), in part by Ministerio de Ciencia e Innovación (Spain) under Grants PID2020-116329GB-C21 and PDC2021-121335-C21, and in part by Universidade de Vigo/CISUG, for open access charge. The review of this article was coordinated by Prof. Ying-Dar Lin. *(Corresponding author: Pablo Fondo-Ferreiro.)*

Pablo Fondo-Ferreiro, Felipe Gil-Castiñeira, Francisco Javier González-Castaño, and David Candal-Ventureira are with the atlanTTic, Information Technologies Group, University of Vigo, 36310 Vigo, Spain (e-mail: pfondo@gti.uvigo.es; xil@gti.uvigo.es; javier@gti.uvigo.es; dcandal@gti.uvigo.es).

Jonathan Rodriguez is with the Instituto de Telecomunicações, 3810-193 Aveiro, Portugal, and also with the University of South Wales, CF37 1DL Pontypridd, U.K. (e-mail: jonathan@av.it.pt).

Antonio J. Morgado is with the University of South Wales, CF37 1DL Pontypridd, U.K. (e-mail: antonio.dasilvamorgado@southwales.ac.uk).

Shahid Mumtaz is with the Department of Applied Informatics Silesian, University of Technology Akademicka, 16 44-100 Gliwice, Poland, and also with the Department of Engineering, Nottingham Trent University, NG1 4FQ Nottingham, U.K. (e-mail: dr.shahid.mumtaz@ieee.org).

Digital Object Identifier 10.1109/TVT.2023.3297017

by the users, but also reducing the overhead introduced by network reconfigurations. To this end, we study the problem of determining at every moment the MEC locations where anchor points are deployed and which anchor point serves each user. Specifically, we focus our study on Vehicle-to-Infrastructure (V2I) scenarios, in which vehicles communicate with MEC applications running on MEC hosts.

In detail, the contributions of this work are:
1) The analytical modeling of the joint problem of the deployment of anchor point Network Functions (NFs) and the assignment of user terminals to anchor points in MEC-enabled scenarios, with the objective of minimizing communications latency while also reducing the overhead caused by network reconfigurations.
2) The proposal of a new heuristic algorithm for solving this problem, and the evaluation of its performance through simulations in a vehicular communications scenario, using publicly available traces of vehicular mobility and base stations' deployments.
3) A comparison of the performance of the newly proposed algorithm with baseline strategies (both centralized and static) and the latency reduction algorithms in [13].

The rest of the article is organized as follows: Section II summarizes related work. Section III defines the problem that is investigated. Section IV presents the proposed solution and the algorithms under evaluation. Section V describes the methodology. Results are presented and discussed in Section VI. Finally, Section VII concludes the article.

## II. RELATED WORK

The MEC paradigm brings computing resources closer to end users. Two main approaches have been previously proposed in the literature to handle user movement between cells while using MEC services: Keeping the running MEC service on the original MEC host [17] or migrating that running service to the MEC host that is associated with the new cell [18]. Even though the first approach is simpler regarding service reconfiguration and migration, it introduces non-negligible overhead in case of frequent handovers, for instance in high-mobility scenarios (e.g. vehicular terminals). Besides, the latency that the users perceive may grow substantially when the served MEC applications do not run on the MEC hosts of the active cells of those users but on the MEC hosts of other cells; and MEC services' migration itself is a costly process that may interrupt these services. For these reasons, some previous works have followed hybrid approaches, by combining MEC service migrations with keeping some services on previous MEC hosts [19].

Among the approaches that propose MEC service migrations, we can identify two main appraches according to the moment when the applications are replicated in the target MEC host: proactive [20], [21] and reactive replication [22]. Some simple proactive approaches consider that the users will move to any of the adjacent cells at some point [20]. More sophisticated proactive strategies perform selective replications to some adjacent cells based on terminal mobility predictions [21]. On the one hand, the main drawback of proactive approaches is the overhead in resource usage due to the deployment of unused instances, which also increases energy consumption. On the other hand, reactive approaches [22] start the migration of a service for a terminal just after it gets attached to the new cell, which minimizes resource usage, but the service may be interrupted while the migration process takes place.

MEC-enhanced vehicular communications have gained attention in the last years [9], [10], [11], [12], [23]. The benefits of MEC in terms of end-to-end latency reduction are showcased in [10], where the authors compare the performance of a conventional cloud-based architecture with a MEC-assisted cellular architecture through simulations of a freeway environment. They report a latency reduction up to an 80% with MEC-aware systems compared to a traditional cloud architecture. The authors of [11] investigate the problem of migrating a service instance from one MEC host to another by following the movement of the vehicles movements to minimize latency. They measure service migration times using Docker containers, and they state that prior knowledge of the trajectories of the vehicles can further reduce service downtime. The work in [9] shows the potential of MEC in C-V2X for cooperative autonomous driving. It presents a system prototype for vehicle groups based on Next Generation Radio Access Network (NG-RAN) and MEC servers providing a High Definition (HD) map service. It also presents two optimization tools based on Artificial Intelligence (AI) to predict the number of users and the network traffic in 5G-V2X cells. The work in [12] focus on the estimation of vehicular mobility to predict cell association changes and proactively deploy services using Virtual Machines (VMs) on the destination MEC host. The mobility predictions are based on a combination of Neural Networks (NN) and Markov chains, leveraging network-based terminal Angle-of-Arrival (AoA) positioning supported by Multiple-Input Multiple-Output (MIMO) technology. They use an online Lyapunov algorithm to determine where and when VMs are replicated. Simulation results show a reduction of energy consumption by 50% compared to full proactive replication strategies, with a bounded risk of continuity loss in computation tasks. The authors of [24] also study latency-critical services in vehicular networks supported by MEC platforms. They consider a three-tiered system for computation task offloading, where the tasks can be executed at vehicle, MEC or backhaul network levels. They apply a Reinforcement Learning (RL) algorithm for making offloading decisions, which results in improved latency and energy consumption, compared with static scenarios. In [23] the Follow Me edge-Cloud (FMeC) architecture for V2I communications is proposed. It ensures that the vehicles are always connected to the closest MEC host by reactively responding to user mobility. In [25], the authors explore dynamic MEC provision of video delivery services to users inside a high speed train. They implement a proof of concept based on a 5G network architecture that dynamically and proactively populates video chunks in MEC hosts based on mobility predictions for improving cache hit ratios. The work in [26] studies joint offloading and resource allocation decisions in vehicular fog-edge scenarios. It formulates the offloading of computing tasks involving vehicles, Road-Side Units (RSUs) and MEC servers as a Stackelberg game and propose incentive mechanisms to motivate vehicles to share their idle resources.

In our previous work we proposed a dynamic optimization framework for deploying anchor points that minimized the latency perceived by end users [13], but it did not take into account the overhead introduced in the network. Moreover, that previous work did not consider any analytical system model and only focused on the deployment of anchor points, always serving the users from the closest anchor point. In this work, we enhace that framework to also select the serving anchor point for each user with the joint objective of minimizing both the latency and the network overhead.

Overall, most prior work on dynamic MEC has only analyzed the cost of migrating computing resources (i.e. VMs or containers) while neglecting network reconfiguration costs. In this work, however, we take into account the overheads incurred both when deploying anchor point NFs at edge locations and when migrating user contexts to the desired locations using our SDN-based solution for transparent SSC. This solution consists in deploying anchor point NFs at edge locations, by taking advantage of virtualization technologies, and migrating the internal context of the anchor point to the replicated edge instances. Then, SDN switches at the edge are reconfigured to forward the traffic to the edge anchor point. The solution is valid both for 4G and 5G networks, in which the anchor point NF respectively corresponds to the Serving/PDN Gateway (S/P-GW) and the User Plane Function (UPF). The internal workings of the solution are described in [15].

We first formulate the problem of joint anchor point deployment and terminal assignment, by considering the trade-off between the latency perceived by the users and the overhead introduced in the network, as an Integer Linear Programming (ILP) multi-objective optimization model. Then, we propose a novel heuristic greedy algorithm to solve it efficiently. Finally, we compare the performance of this algorithm with baseline strategies (centralized and static) and other state of the art alternatives, such as the latency reduction approaches in [13].

## III. PROBLEM STATEMENT

We consider a MEC-assisted vehicular communications scenario with low latency and low throughput requirements. This type of communication corresponds to safety-related applications such as collision avoidance [8], among others. This is a latency-critical scenario in which the contribution of communication latency to overall task offloading time may be significant. Safety applications can be pre-deployed on the MEC hosts involved, so the deployment time for the MEC applications can be neglected. Moreover, we consider that the MEC applications do not execute complex calculations but mainly gather the data sent by the vehicles, and thus processing time of the MEC applications can also be neglected.

In this context, the vehicles communicate through a cellular network with an application that can be deployed at the core network or in a MEC host. This application can be migrated to a different location during its execution. As vehicles move, they get associated to new base stations and therefore the latency of the communications with the application instance varies. In order to ensure service continuity, we leverage our solution for seamless anchor point migrations in dynamic MEC environments [15]. As previously mentioned, the solution allows deploying anchor point NFs in MEC hosts and reconfiguring the SDN network to forward the traffic of each individual vehicle to the desired location. Basically, the SDN switches at the network edge redirect the traffic of each User Equipment (UE) to the desired anchor point.

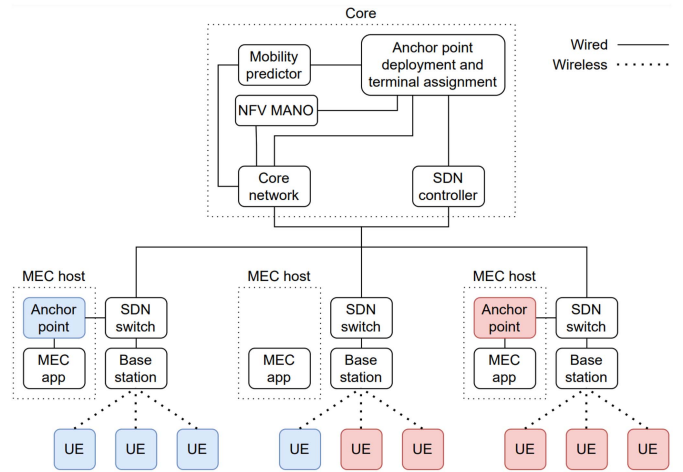

Fig. 1. System architecture for the problem considered (UE assignments to anchor points are marked in red or blue depending on the anchor point).

In this scenario, we seek to dynamically reconfigure the network departing from current vehicle locations and UE assignments to anchor points, to minimize communications latency and network overhead. In detail, our proposal predicts vehicle mobility and then decides the deployment and removal of anchor point NFs in the MEC hosts as well as the assignment of UEs to anchor point locations. We remark that reducing the number of anchor points is important to reduce the computing overhead in terms of resource usage and energy consumption caused by unused instances, as previously said.

We remark that the main difference with respect to existing works on MEC application deployment lies in the consideration of the overheads introduced by a practical SDN-based solution for session and service continuity. The analysis and minimization of these overheads has been neglected in previous works and therefore existing solutions cannot be directly applied to the problem.

Fig. 1 illustrates the considered system architecture. Anchor point network functions are deployed in MEC hosts, and UEs are assigned to them according to the decisions of the anchor point deployment and terminal assignment module.

### A. System Model

We consider a time-slotted system model. At the beginning of each time slot, given the current state, the network intelligence determines the edge locations where the anchor points must be deployed and which edge location should be used for serving each user. The current state includes current vehicle locations (i.e., which base station they are connected to), the currently deployed anchor points and the current assignments of anchor points to user terminals. The decisions consist of anchor points' deployments at edge locations where no other active anchor points were present beforehand, removal of anchor points from edge locations that are no longer used, and reconfiguration of the SDN switches of the network for diverting the traffic of each user to the desired anchor point. Changes are only made at the beginning of each time slot and they remain valid for the duration of that slot.

TABLE I
INPUT DATA OF THE PROBLEM

| | |
|---|---|
| $\mathcal{N}$ | Set of nodes in the network topology. |
| $\mathcal{L}$ | Set of links in the network topology (with link latency weights $w_{ij}$). |
| $\mathcal{E}$ | Subset of network nodes $\mathcal{E} \subset \mathcal{N}$ that are base stations with edge nodes. |
| $c$ | Node corresponding to the core anchor point. |
| $\mathcal{V}$ | Set of vehicles in the scenario. |
| $x_{ij}$ | Binary variable that is equal to 1 if vehicle $i$ is currently connected to base station $j$ and 0 otherwise. |
| $\hat{x}_{ij}$ | Binary variable that is equal to 1 if vehicle $i$ is predicted to be currently connected to base station $j$ and 0 otherwise. |
| $y_i$ | Binary variable that is equal to 1 if an anchor point is currently deployed at edge location $i$ and 0 otherwise. |
| $z_{ij}$ | Binary variable that is equal to 1 if vehicle $i$ is currently assigned to the anchor point at edge location $j$. |
| $a$ | Cost for deploying an anchor point. |
| $b$ | Cost for removing an anchor point. |
| $o_{ij}$ | Cost for relocating the communications of a vehicle from the anchor point at location $i$ to the anchor point at location $j$. |
| $N_{\text{anchor\_points}}$ | Number of anchor points to be deployed. |

TABLE II
OUTPUT DECISION VARIABLES

| | |
|---|---|
| $y'_i$ | Binary variable that is equal to 1 if an anchor point is scheduled to be deployed at edge location $i$ and 0 otherwise. |
| $z'_{ij}$ | Binary variable that is equal to 1 if vehicle $i$ is scheduled to be served by the anchor point at edge location $j$. |

We formulate the assignment problem at the beginning of each time slot as an NP-hard ILP.

*Input data of the problem:* We model the network topology as an undirected weighted graph $\mathcal{G} = (\mathcal{N}, \mathcal{L})$, where $\mathcal{N}$ is a set of $N$ nodes that represent the forwarding devices in the network topology and $\mathcal{L}$ is a set of $L$ links that interconnect the nodes in $\mathcal{N}$. Each link in $\mathcal{L}$ from node $i$ to node $j$ is weighted with the latency $w_{ij}$ of the link. The core anchor point will be the root node $c \in \mathcal{N}$ of the hierarchical network topology. A subset $\mathcal{E} \subset \mathcal{N}$, represents the $E$ edge locations where anchor points can be placed. We consider that base stations are co-located with edge locations. From the link latency values $w_{ij}$ we can derive $l_{ij}$, the latency value from node $i$ to node $j$ for $i, j \in \mathcal{N}$, as the sum of the link weights in the shortest path $\text{sp}(i,j)$ from node $i$ to node $j$ (1).

$$l_{ij} = \sum_{k \to l \in \text{sp}(i,j)} w_{kl} \quad (1)$$

Set $\mathcal{V}$ represents the $V$ vehicles in the scenario. Without loss of generality, we only consider vehicles demanding low-latency services (any other services are irrelevant to the problem in this work, since they may be served through any anchor point, such as the core anchor point itself). Current vehicle locations are input data for our problem. We define the current vehicle connection matrix $\mathbf{X} = (x_{ij})$ as an $V \times E$ binary matrix, where $x_{ij}$ is 1 if vehicle $i \in \mathcal{V}$ is connected to the base station at edge location $j \in \mathcal{E}$ and 0 otherwise. Since a vehicle can only be attached to a single base station (note that by "assignment" we refer to the location of the anchor point that is serving a vehicle terminal and by "connection" to the location of the base station to which the terminal is attached), input vehicle connections satisfy:

$$\sum_{j \in \mathcal{E}} x_{ij} = 1, \quad \forall i \in \mathcal{V}. \quad (2)$$

To represent a more realistic scenario, our algorithm does not directly work with the current real vehicle connection matrix ($\mathbf{X}$), but with a predicted vehicle connection matrix based on previous vehicle positions, given by $\hat{\mathbf{X}} = (\hat{x}_{ij})$, which is analogous to $\mathbf{X}$, but using predicted values.

Other input variables are $a$, the cost for deploying an anchor point; $b$, the cost for removing an anchor point; and $o_{ij}$ the cost for relocating the communications of a vehicle between the anchor points at locations $i$ and $j$, which is the control-plane information overhead introduced in the network for this reason. Table I summarizes the input data of the problem.

*Output decision variables:* The vector of the next deployments is a binary vector $\mathbf{y}' = (y'_i)$ of dimension $E$ such that $y'_i$ is 1 if an anchor point is scheduled to be deployed at edge location $i \in \mathcal{E}$ and 0 otherwise. The matrix of the next vehicle assignments is a $V \times (E + 1)$ binary matrix $\mathbf{Z}' = (z'_{ij})$ such that $z'_{ij}$ is 1 if vehicle $i \in \mathcal{V}$ is scheduled to be served by the anchor point at location $j \in \mathcal{E} \cup \{c\}$.

Note that the output decision variables of one slot ($\mathbf{y}'$ and $\mathbf{Z}'$) become the input data for the following slot ($\mathbf{y}$ and $\mathbf{Z}$). Initially, all vehicles are served by the core anchor point when joining the network (i.e., $\mathbf{Z} = \mathbf{0}$) and no anchor point is deployed at any edge location (i.e., $\mathbf{y} = \mathbf{0}$), as in a realistic scenario. The same applies to new vehicles joining the network: the first time slot they appear in, they are served by the core anchor point. Table II summarizes the output decision variables.

*Objective functions:*
1) Vehicle latency ($90^{th}$-percentile):

$$f_1 = P_{90\%}\left(\left\{\sum_{j \in \mathcal{E}, k \in \mathcal{E} \cup \{c\}} x_{ij} z'_{ik} l_{jk}\right\}, i \in \mathcal{V}\right), \quad (3)$$

where $P_{90\%}$ picks the latency marking the $90^{th}$ percentile.

2) Deployment overhead:

$$f_2 = \sum_{i \in \mathcal{E}} [\max(y'_i - y_i, 0) \cdot a + \max(y_i - y'_i, 0) \cdot b]. \quad (4)$$

3) Control-plane reassignment overhead introduced in the network:

$$f_3 = \sum_{i \in \mathcal{V}} \sum_{j \in \mathcal{E} \cup \{c\}} \sum_{k \in \mathcal{E} \cup \{c\}} z_{ij} z'_{ik} o_{jk}. \quad (5)$$

From expressions (3)–(5), the multi-objective optimization goal is defined as:

$$\text{Minimize } (f_1, f_2, f_3). \quad (6)$$

This multi-objective optimization is converted to a single-objective optimization with a linear weighted scalarization [27] of normalized objective function components to $[0, 1]$, using weights $\alpha_1, \alpha_2, \alpha_3$ such that $\sum_{i=1}^{3} \alpha_i = 1$. The resulting objective function is:

$$\text{Minimize } \left( \frac{\alpha_1 f_1}{\max(f_1)} + \frac{\alpha_2 f_2}{\max(f_2)} + \frac{\alpha_3 f_3}{\max(f_3)} \right), \quad (7)$$

where $\max(f_1) = \max_{i,j \in \mathcal{N}}(l_{ij})$ is the maximum latency that can be perceived by any vehicle, which is given by the graph diameter (i.e., the maximum distance between any pair of vertices); $\max(f_2) = N \cdot \max(a, b)$ is the maximum deployment overhead; and $\max(f_3) = V \cdot \max(o_{ij})$ is the maximum control-plane reassignment overhead that can be introduced in the network. Note that the selection of $\alpha_i$ values allows for modulating the trade-off between the relevance of the different objective functions in the multi-objective optimization problem.

Note also that we have considered the overhead as part of the objective function because this way we do not need to set an arbitrary bound on the maximum network overhead allowed. Instead, we let the algorithm find a minimal amount of network overhead while also minimizing latency.

*Constraints:*

1) Assignment unicity: a vehicle can only be assigned to a single anchor point.

$$\sum_{j \in \mathcal{E} \cup \{c\}} z'_{ij} = 1, \quad \forall i \in \mathcal{V}. \quad (8)$$

2) Deployment of required anchor points: anchor points should be deployed at every edge location with assigned vehicles.

$$y'_j = 1, \forall j \in \mathcal{E} \mid \sum_{i \in \mathcal{V}} z'_{ij} > 0. \quad (9)$$

3) Resource usage: The number of anchor points that are scheduled to be deployed must be $N_{\text{anchor\_points}}$.

$$\sum_{i \in \mathcal{E}} y'_i = N_{\text{anchor\_points}}. \quad (10)$$

Finally, note that our system model can be easily extended to consider network link bandwidths or MEC node capacities by adding vehicle throughput as problem input data and the corresponding constraints to the model. In a real scenario, the actual measurements of vehicle throughput can be directly obtained from the flow counters of the SDN switches. However, we did not include them in our model because they are not relevant to our problem statement, since we are only considering communications with low throughput requirements, which do not saturate network links nor nodes.

We could have also considered computing resources on MEC hosts with appropriate constraints in the system model. However, we are approaching the problem from a network layer perspective while abstracting computing resources, which could be handled by complementary future work.

## IV. PROPOSED SOLUTION

The architecture of the proposed solution extends the dynamic optimization framework presented in [13] to allow selecting the serving MEC location for each UE. Let us recall that the architecture includes SDN switches at the gNBs. These switches are configured by the SDN controller to steer the traffic of each UE to the desired anchor point. An anchor point deployment and terminal assignment algorithm solves the problem in Section III to determine the locations where anchor points will be deployed and assign UEs to these anchor points. The deployment decision is sent to the NFV Management and Orchestration (MANO) platform, which deploys the corresponding anchor point VNFs at the requested MEC hosts. The resulting assignments of UEs to anchor points are also communicated to the SDN controller, which reconfigures the network for relocating the UEs to the intended anchor points. Taking this into consideration, in practice the algorithm should be deployed at the operator's core network for reducing the signaling to interact with the NFV MANO and the SDN controller. Note that these two entities also provide the input data to the algorithm (Table I), except for the predicted vehicle connection matrix, which is calculated using a Long Short-Term Memory (LSTM) [28], following the same approach of similar works [29]. The network reconfiguration for relocating a UE involves the replication of the UE context from the previous anchor point to the new one and also the reconfiguration of SDN flow rules at the corresponding edge switches. The outputs of the algorithm are the decision variables in Table II, which identify the MEC hosts where the anchor points are deployed and the assignment of UEs to anchor points at the current time slot.

We first present the latency minimization algorithms for vehicular communication scenarios described in [13] for the sake of clarity. These algorithms seek to minimize the latency (i.e., $f_1$) for a given level of resource usage ($N_{\text{anchor\_points}}$, number of anchor points deployed), without considering the overhead introduced in the network. Then, we present our proposed novel *overhead-aware greedy average* heuristic to solve the multi-objective optimization problem (7).

### A. Latency-Minimization Algorithms

The latency-minimization algorithms in the following list operate as follows: first, the anchor point deployment algorithm determines the subset of edge nodes to deploy anchor points for a given resource usage. Then, each UE is assigned to the closest anchor point.

- *Centralized:* This is a first baseline strategy that only considers the centralized static deployment of a single anchor point in the core network. Therefore, it does not involve any re-deployment of anchor points nor any UE re-assignments.
- *Static K-means:* This is a second baseline strategy, which ignores the distribution of the UEs and performs a static deployment of anchor points at fixed locations that are determined when the network is built. It first clusters all the edge locations in the network with the K-means algorithm. Then, the edge site that is closest to each cluster center in Euclidean distance is chosen for deploying an anchor point. Therefore, this strategy does not re-deploy anchor points. However, as the UEs move, their closest anchor point may change, and, as a result, anchor point assignments to UEs may change.
- *Random:* This is a third baseline strategy that selects at random the subset of edge sites for the deployment of anchor points, uniformly across all available edge sites.
- *Greedy percentile:* This heuristic follows a greedy strategy that determines edge sites iteratively. At each iteration, the algorithm selects the edge site that would result in the lower $90^{th}$-percentile latency. In case of multiple sites providing the same value, the algorithm selects the site that would provide the lowest average latency.
- *Greedy average:* Heuristic based on an iterative greedy strategy that chooses the edge site providing the lowest average latency.
- *K-means:* This strategy is based on a proposal in [30]. At each time slot, active edge sites (i.e., those with UEs attached to them) are clustered with a K-means algorithm. Then, anchor points are deployed at the closest edge sites using the Euclidean distance.
- *K-means greedy average:* At each time slot, active edge sites are clustered with a K-means algorithm. Then, each cluster is independently considered and one anchor point is deployed in each cluster. The edge site for deploying the anchor point in each cluster is selected to minimize the latency perceived by the UEs that are attached to the nodes of the cluster.
- *Modularity greedy average:* At each time slot, the active edge sites are clustered with a modularity maximization-based strategy using the Louvain algorithm [31]. Then, each cluster is independently considered and one anchor point is deployed in each cluster. As in the previous case, the edge site for deploying the anchor point in each cluster is selected to minimize the latency perceived by the UEs that are attached to the nodes of the cluster.

In the clustering algorithms, we set the number of clusters to match the number of anchor points to be deployed. We remark that, if a different MEC deployment strategy is used (e.g., with one MEC host serving multiple base stations), the proposed strategies can be generalized by introducing the corresponding constraints (e.g., by only deploying anchor points at allowed MEC locations).

### B. Overhead-Aware Algorithm

We propose a latency-minimization algorithm to not only minimize the latency, but also to reduce all the objective functions defined in Section III, also including the deployment overhead ($f_2$) and the control-plane reassignment overhead introduced in the network ($f_3$). In detail, the algorithm works as follows:

*Overhead-aware greedy average:* Heuristic algorithm based on a greedy strategy that iteratively chooses the edge site that results in the lowest objective function value in (7). In the internal calculations of the algorithms, we consider the average latency rather than the $90^{th}$-percentile in $f_1$ for reducing the computational complexity. The difference by considering the average rather than the $90^{th}$-percentile is negligible in terms of the actual latency values achieved, as we can observe in the comparison between the greedy percentile and the greedy average algorithms in Section VI. As previously said, UEs are not necessarily assigned to the closest anchor point, but to the anchor point that minimizes the objective function (e.g., in some situations, the algorithm chooses to maintain users in their previous location to reduce the overhead introduced in the network). The proposed algorithm is detailed in Algorithm 1. The outer loop ensures that $N_{\text{anchor\_points}}$ anchor points are selected. At each iteration, one anchor point is incrementally chosen. The anchor point producing the largest descent of the objective function is selected to be deployed. In the inner loop, we individually consider whether each vehicle should be served by the current anchor point or stay assigned to the previously selected anchor point. For speedup purposes, rather than evaluating each vehicle individually, the algorithm jointly evaluates all vehicles assigned to the same anchor point in each base station. In this way, the computational complexity of the algorithm is only $\mathcal{O}(N_{\text{anchor\_points}}^2 \cdot E^2)$ instead of $\mathcal{O}(N_{\text{anchor\_points}} \cdot E \cdot V)$, so that it is independent of the number of vehicles ($V$), which can be very large. Note that this change does not introduce any difference in the output produced by the algorithm, since the vehicles connected to the same base station that were previously assigned to the same anchor point will always be assigned to the same anchor point, even if they are individually evaluated.

## V. METHODOLOGY

This section describes the methodology we have followed to evaluate and compare the algorithms for anchor point deployment and terminal assignment in a MEC-enabled vehicular communications scenario.

We begin with a description of the datasets that we have used for our experiments, and then we present the main metrics we used to evaluate the performance of the different strategies.

### A. Dataset

The algorithms have been evaluated using two publicly available urban mobility datasets [32], respectively, containing traffic traces of realistic car trips and the location of real base stations in the same urban environment.

This environment corresponds to an area of $400 \, \text{km}^2$ of Cologne, Germany. The vehicular mobility dataset has a size

**Algorithm 1:** *Overhead-Aware Greedy Average* Algorithm.

**Input**: Nodes $\{\mathcal{N}, \mathcal{E}, c\}$, Links $\{\mathcal{L}, w_{ij}\}$, Vehicles $\mathcal{V}$, Current vehicle connections $\{x_{ij}\}$, Current anchor point deployments $\{y_i\}$, Current vehicle assignments $\{z_{ij}\}$, Deployment cost $a$, Removal cost $b$, Relocation costs $\{o_{ij}\}$.
**Output**: Scheduled anchor point deployments $\{y'_i\}$, Scheduled vehicle assignments $\{z'_{ij}\}$.
**Parameters**: $\alpha_1, \alpha_2, \alpha_3, N_{\text{anchor\_points}}$.

$l_{ij} \leftarrow \sum_{k \leftrightarrow l \in \text{sp}(i,j)} w_{k,l}$ for all $i, j$;
$y'_i \leftarrow 0$ for all $i$;
$z'_{ij} \leftarrow 0$ for all $i, j$;
$z''_{ij} \leftarrow 0$ for all $i, j$;
**repeat** $N_{\text{anchor\_points}}$ **times**
$\quad e' \leftarrow -1$;
$\quad f' \leftarrow +\infty$;
$\quad$**for** *each $e$ in $\mathcal{E}$ such that $y'_i = 0$* **do**
$\quad\quad$// Check tentative deployment of anchor point at location $e$
$\quad\quad y'_e \leftarrow 1$;
$\quad\quad f_2 \leftarrow \sum_{i \in \mathcal{E}}[\max(y'_i - y_i, 0) \cdot a + \max(y_i - y'_i, 0) \cdot b]$;
$\quad\quad f_{2,\text{norm}} \leftarrow \frac{\alpha_2 f_2}{N \cdot \max(a,b)}$;
$\quad\quad$**for** *each $v$ in $\mathcal{V}$* **do**
$\quad\quad\quad$// Check served vehicle $v$ from $e$
$\quad\quad\quad f_{1,\text{keep}} \leftarrow \frac{\sum_{j \in \mathcal{E}, k \in \mathcal{E} \cup \{c\}} x_{vj} \cdot z'_{vk} \cdot l_{jk}}{V}$;
$\quad\quad\quad f_{3,\text{keep}} \leftarrow \sum_{j \in \mathcal{E} \cup \{c\}} \sum_{k \in \mathcal{E} \cup \{c\}} z_{vj} z'_{vk} o_{jk}$;
$\quad\quad\quad f_{1,\text{relocate}} \leftarrow \frac{\sum_{j \in \mathcal{E}} x_{vj} \cdot l_{je}}{V}$;
$\quad\quad\quad f_{3,\text{keep}} \leftarrow \sum_{j \in \mathcal{E} \cup \{c\}} z_{vj} o_{je}$;
$\quad\quad\quad f_{13,\text{keep}} \leftarrow \frac{\alpha_1 f_{1,\text{keep}}}{\max_{\forall i,j \in \mathcal{N}}(l_{ij})} + \frac{\alpha_3 f_{3,\text{keep}}}{V \cdot \max(o_{ij})}$;
$\quad\quad\quad f_{13,\text{relocate}} \leftarrow \frac{\alpha_1 f_{1,\text{relocate}}}{\max_{\forall i,j \in \mathcal{N}}(l_{ij})} + \frac{\alpha_3 f_{3,\text{relocate}}}{V \cdot \max(o_{ij})}$;
$\quad\quad\quad f \leftarrow f_{2,\text{norm}} + \min(f_{13,\text{keep}}, f_{13,\text{relocate}})$;
$\quad\quad\quad$**if** $f_{13,relocate} < f_{13,keep}$ **then**
$\quad\quad\quad\quad z''_{vj} \leftarrow 1$ if $j = e$, otherwise 0;
$\quad\quad\quad$**end**
$\quad\quad$**end**
$\quad\quad$**if** $f < f'$ **then**
$\quad\quad\quad f' \leftarrow f$;
$\quad\quad\quad e' \leftarrow e$;
$\quad\quad$**end**
$\quad\quad y'_e \leftarrow 0$;
$\quad$**end**
$\quad$// Apply best deployment and assignment
$\quad y'_{e'} \leftarrow 1$;
$\quad z'_{ij} \leftarrow z''_{ij}$ for all $i, j$;
**end**
**return** $\{\mathbf{y}', \mathbf{z}'\}$

location of the vehicle in Cartesian coordinates and its speed[1]. The base station deployment dataset contains the Cartesian coordinates of 247 base stations retrieved from public German databases[2]. This dataset contains information about the real deployment of base stations belonging to all operators in the area.

To create the network topology graph for our experiments, we assumed that each base station is directly connected to other base station in their neighborhood, so that a single connected graph results. According to our problem formulation, MEC sites are co-located with base stations, and therefore operators can deploy anchor points and execute MEC applications on them.

Each base station in the deployment dataset was considered a node in our graph. Then, we employed the following methodology to create undirected links between the nodes, in the same way as in [13]:

- Initially, we connect each base station to the base station in its neighborhood. We considered that two base stations belong to the same neighborhood if their Euclidean distance is less than $DTHRESHOLD$.
- Then, we take the set of connected components resulting from the previous step and incrementally build a connected graph by iteratively joining the two largest components. To do this, we set a link between the closest two nodes (i.e., those separated by the shortest Euclidean distance), such that one node belongs to the largest component and the other belongs to the second largest component.

This procedure creates an undirected connected graph. In addition, we also set the weight of each link to 1 (i.e., $w_{ij} = 1$), which is equivalent to an unweighted graph.

The neighborhood threshold $DTHRESHOLD$ was set to 500 meters in our experiments, resulting in a total of 293 edges. The minimum graph distance (i.e. in number of hops) between every pair of nodes is normally distributed with an average value of 18 hops. Note that the neighborhood threshold is not a parameter of our proposed solution, but just an auxiliar parameter used during the dataset preprocessing step to generate a graph between the base stations. In our experiments, after analyzing the dataset, we empirically chose the value of 500 meters to generate links between nearby base stations.

Since our model is time-slotted, we jointly consider the entries of the vehicular mobility dataset whose timestamps lie within the same 5-second slot. For attaching UEs to base stations, we followed the same methodology as in related works in the literature [12]:

- If the UE was not present in the previous time slot, we attach it to the base station that minimizes the path loss (i.e., the closest base station). The path loss is calculated using the formula for Non-Line-Of-Sight (NLOS) urban environments [33]. Specifically:

$$\overline{PL}(d)[dB] = \alpha + 10 \cdot \beta \cdot log(d) + X_\sigma, \quad (11)$$

of $\sim$20 GB and has 354 million entries describing more than 700 000 synthetic car trips during a 24-hour interval of a typical workday. This realistic synthetic dataset captures both the macroscopic and microscopic dynamics of road traffic in an urban area. Each entry consists of a simulation timestamp, the

---
[1]The vehicular mobility dataset is publicly available at http://kolntrace.project.citi-lab.fr/koln.tr.bz2.
[2]The base station deployment dataset is publicly available at http://kolntrace.project.citi-lab.fr/koln_bs-deployment-D1_fixed.log.

where $\alpha$ is the floating intercept and $\beta$ is the slope, both computed as least-squares fits; $d$ corresponds to the Euclidean distance between the UE and the base station; and $X_\sigma$ models the effect of shadowing as a zero-mean Gaussian random variable with variance $\sigma^2$. According to [33], we have set $\alpha = 46.61, \beta = 3.63$ and $\sigma = 9.83$ dB.
- If the UE was already present in the previous time slot, we compare the path loss between the vehicle UE and the serving base station with the minimum path loss with any other base station. A hysteresis margin $\epsilon$ was considered to trigger the handover to the new base station. If the path loss improvement is less than $\epsilon$, the UE remains attached to the previous base station during the current time slot. We followed the recommendation in [34] to set $\epsilon = 2$ dB.

The same procedure has been applied to determine the predicted connection matrix ($\hat{\mathbf{X}}$), but using predicted positions instead of the real ones. The predicted positions have been computed using an LSTM network composed of two stacked LSTM cells with 50 hidden units each. This network has been trained during 10 epochs and a batch size of 1000 samples. We applied a 20%–80% split to the dataset to obtain the training and testing sets, respectively. The LSTM network has an RMSE prediction value of 46.19 compared with an RMSE of 161.61 in the case of a baseline naive algorithm (using the last value as the prediction).

*B. Evaluation Metrics*

We have evaluated the different objective functions defined in the problem statement (namely the latency perceived by the UEs $f_1$, the deployment overhead $f_2$ and the control-plane reassignment overhead $f_3$) for different levels of resource usage $N_{\text{anchor\_points}}$. In addition, we also checked the running times of the algorithm.
- *Latency perceived by the UEs:* It measures the communications latency from the UEs to their serving anchor points as given by (3). We calculate each UE latency as the sum of the link latency values in the communication path between the UE and its serving anchor point. The latency of each link in the backhaul network will strongly depend on the technology of the link. In our evaluations, we are assuming that all links between base stations are equal and thus we use unitary latency links (i.e., $w_{ij} = 1$). In this case, the communication latency is equivalent to the number of hops between the user and the serving anchor point. We consider the $90^{th}$-percentile values perceived by all the UEs as an aggregate.
- *Deployment overhead:* It measures the overhead introduced by removing and deploying anchor points. It is given by $f_2$ in (4). To calculate it we set $a = 1$ and $b = 0.1$ for reflecting the relative magnitudes of the times required for deploying and removing anchor points, where the latter is usually lower.
- *Control-plane reassignment overhead:* It measures the overhead introduced by changing the serving anchor point for the UEs. It is given by $f_3$ in (5). To calculate it, we set the cost for relocating the communications of a UE as the number of links in the shortest path from the location of the previous anchor point to the location of the new anchor point. Formally, we thus set $o_{ij} = |\{k \leftarrow l \in \text{sp}(i,j)\}|$.
- *Algorithm running time:* It is the elapsed time for the execution of the anchor point deployment and terminal assignment algorithm, since the algorithm receives the input data until it generates the corresponding output. It is related to the computational complexity of the decision algorithm in each time slot.

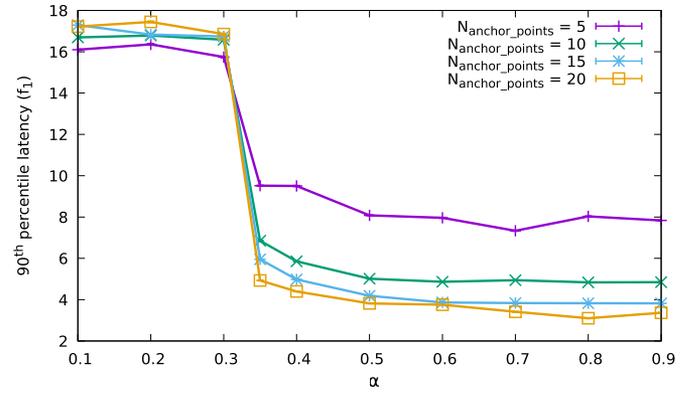

Fig. 2. $90^{th}$-percentile latency perceived by the UEs ($f_1$, number of hops) as $\alpha$ increases for different levels of resource usage using the overhead-aware greedy average algorithm.

## VI. RESULTS AND DISCUSSION

We first evaluated the proposed overhead-aware greedy average algorithm for different values of $\alpha$ in the simulated urban vehicular mobility scenario using the previously described datasets. The code has been written in Python and is publicly available in [35] under an open-source license, for the sake of reproducibility of our results. We have used the PyPy Python implementation [36] as running environment, executed on a Intel Core i9-9900 K CPU @ 3.60 GHz desktop computer.

The trade-off between the different objective functions can be fine-tuned by choosing the $\alpha_1$, $\alpha_2$ and $\alpha_3$ parameters of the scalarization of the model. The first experiments studied the effects of the selection of $\alpha_i$ values in the different objective functions. In these experiments we set $\alpha_1 = \alpha$ and $\alpha_2 = \alpha_3 = \frac{1-\alpha}{2}$ for $\alpha \in (0,1)$ in order to give the same weight to both overheads in the objective function. Then, we evaluated the performance of the overhead-aware greedy average algorithm for $\alpha \in [0.1, 0.9]$ for different resource usages $N_{\text{anchor\_points}} = \{5, 10, 15, 20\}$.

Figs. 2, 3, 4, and 5 show the $90^{th}$-percentile latency perceived by the UEs ($f_1$), deployment overhead ($f_2$), control-plane reassignment latency ($f_3$) and algorithm running time, respectively, for increasing values of $\alpha$ and different levels of resource usage, by applying the overhead-aware greedy average algorithm. As we could expect, increasing $\alpha$ results in lower latency but higher overheads. However, we can distinguish two regions in the figures: for $\alpha$ less than 0.35 the overhead is negligible but there is no latency minimization at all; and for $\alpha$ higher than 0.35, the latency is considerably lower and keeps slightly decreasing with increasing $\alpha$ values, whereas the overheads

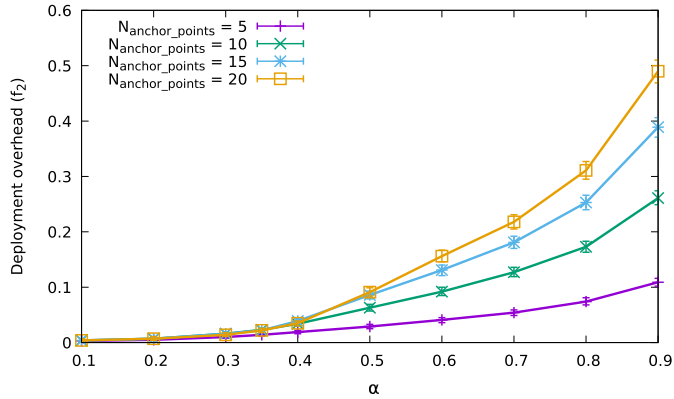

Fig. 3. Deployment overhead ($f_2$, overhead in terms of a combination of the cost of deploying and removing an anchor point) as $\alpha$ increases for different levels of resource usage using the overhead-aware greedy average algorithm.

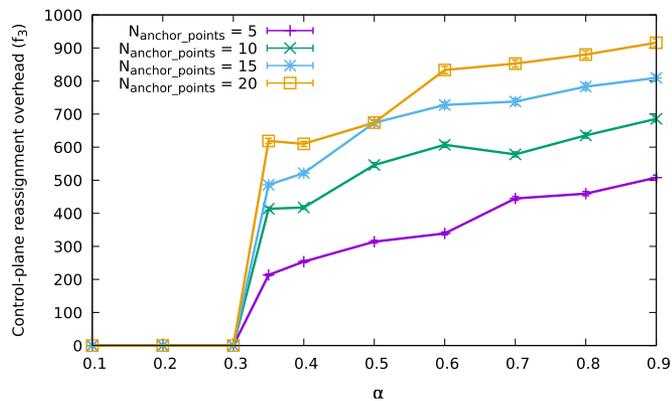

Fig. 4. Control-plane reassignment overhead ($f_3$, overhead in terms of the cost for relocating vehicle communications) as $\alpha$ increases for different levels of resource usage using the overhead-aware greedy average algorithm.

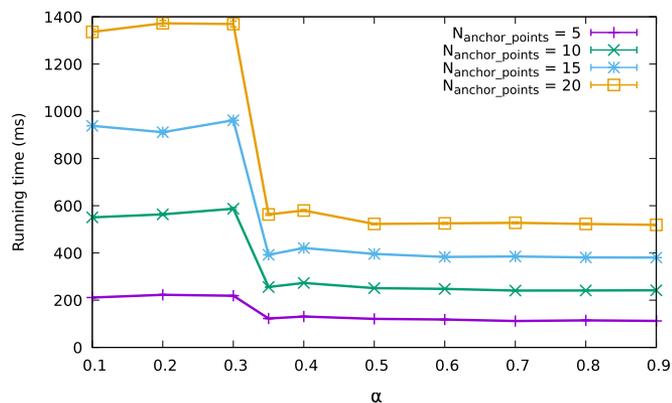

Fig. 5. Average algorithm running time as $\alpha$ increases for different levels of resource usage using the overhead-aware greedy average algorithm.

correspondingly increase. Interestingly, the two regions lead to substantially distinct running times and reveal a difference in the actual computational complexity of the algorithm. In the first region they seem to grow quadratically with resource usage, while in the second region they grow almost linearly.

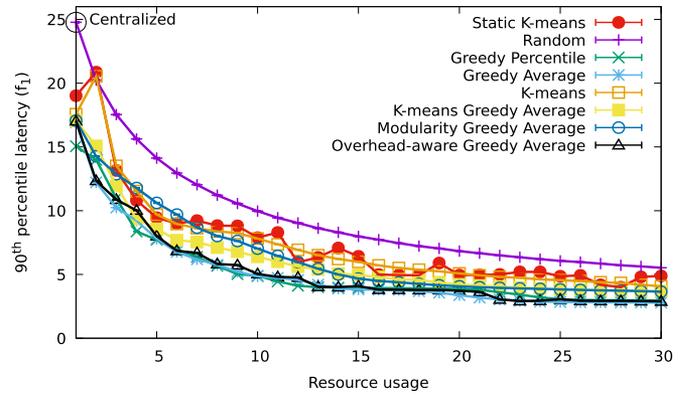

Fig. 6. $90^{th}$-percentile latency perceived by the UEs ($f_1$, number of hops) as resource usage level ($N_{\text{anchor\_points}}$, number of anchor points deployed) increases.

Overall, the first region is not valid for latency minimization, since the algorithm degenerates into choosing the trivial solution of maintaining the first deployment and terminal assignment throughout the whole simulation to avoid any overheads. The inflection point at $\alpha = 0.35$ causes an abrupt change in the behavior of the algorithm, which reduces considerably the latency perceived, while gradually increasing the overheads. Consequently, $\alpha$ should be set in this second region to actually consider latency into the minimization objective. Interestingly, the reduction in the latency is small for increasing $\alpha$ values, while the increase in the overheads is relatively higher. This suggests that values of $\alpha$ around 0.5 achieve a satisfactory trade-off for jointly minimizing the latency and the overheads. Therefore, this setting was used in the rest of the simulations in this work.

In the following experiments, we compare the proposed overhead-aware greedy average algorithm (with $\alpha = 0.5$) with the latency minimization algorithms discussed in [13].

Fig. 6 depicts the $90^{th}$-percentile latency perceived by the UEs as the resource usage (i.e., the number of anchor points deployed) increased. As an indication of statistical significance, the results include 95% confidence intervals. First, we can notice that an increase in resource usage is directly associated with a reduction in the latency that the UEs perceive. We will reproduce here a thorough discussion of the latency and execution time results for the competing latency-minimization algorithms, since the interested reader can find it in [13]. If we focus on the novel overhead-aware greedy average algorithm proposed in this article, it provides low latency values, in the same range as the greedy average and greedy percentile latency-minimization algorithms.

Fig. 7 shows the average running time for the different algorithms versus the number of anchor points considered. Note that the static strategies (centralized and static K-means) are not considered in this figure because they do not take any decision at every time slot, but instead they maintain the same initial deployment throughout all the time slots of the simulation. We can first observe that the running time increases linearly with the use of resources. A detailed discussion on the execution time of the alternative latency minimization algorithms is also provided in [13] and thus omitted here. Focusing on the proposed overhead-aware greedy average algorithm, it takes a longer execution time than the greedy percentile algorithm, about 800 ms for a resource usage of 30 anchor points, whereas the greedy percentile algorithm takes 500 ms for the same resource usage. The main reason for this behavior seems related to the computational complexity of the worst-case scenario for the overhead-aware greedy average algorithm, which is $\mathcal{O}(N^2_{\text{anchor\_points}} \cdot E^2)$, whereas the greedy percentile algorithm has a computational complexity of $\mathcal{O}(N_{\text{anchor\_points}} \cdot E^3)$. Despite the slight asymptotic reduction of computational complexity, the increase in the constant factors caused by the overhead calculations cannot be neglected. Finally, note that despite the quadratic relationship with resource usage $N_{\text{anchor\_points}}$ in the worst-case scenario, the typical relationship is indeed linear in our experimental results. This is directly related to the fact that the UEs connected to a given base station are previously assigned to a small set of anchor points for $\alpha > 0.35$, as previously discussed. Note that this does not hold for $\alpha$ less than 0.35, because in that case UEs stay assigned to the original anchor points throughout most of the simulation to keep a low $f_3$ overhead.

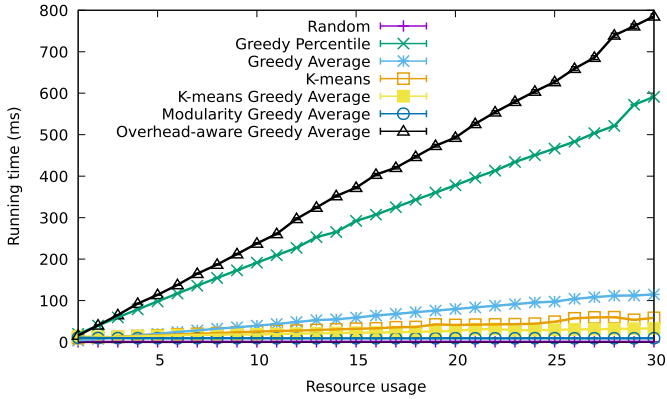

Fig. 7. Average algorithm running time as resource usage level ($N_{\text{anchor\_points}}$, number of anchor points deployed) increases.

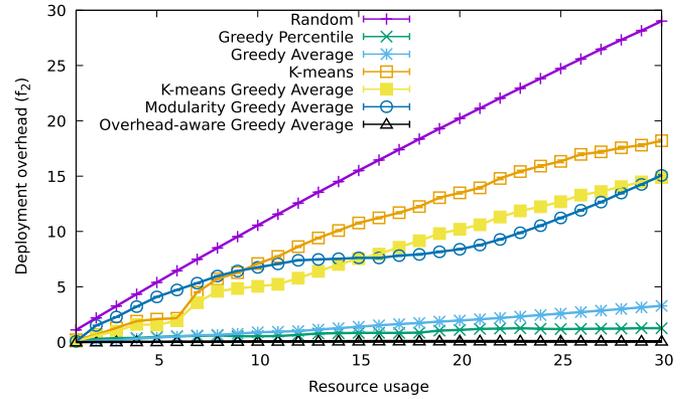

Fig. 8. Deployment overhead ($f_2$, overhead in terms of a combination of the costs of deploying and removing an anchor point) as resource usage level ($N_{\text{anchor\_points}}$, number of anchor points deployed) increases.

It is important to have in mind that the time to deploy the required anchor points and reassign the vehicles' UEs to the desired locations should be shorter than the slot duration. This time comprises the deployment of the anchor point, the execution of the assignment algorithm and also the time to deploy the anchor points and reconfigure the SDN network for diverting the traffic of each UE to the new anchor point. As shown by our results in [13], the deployment of anchor points takes in the order of one second, while network reconfiguration take is in the order of tens of milliseconds with our SDN-based mechanism for transparent SSC. Taking this into consideration, the the time slot should not last for less than a couple of seconds for ensuring system stability. Indeed, this also highlights the relevance of reducing the deployment overhead ($f_2$) and control-plane reassignment overhead ($f_3$) metrics, to avoid unnecessary overhead.

Fig. 8 shows the deployment overhead $f_2$ in (4) as the resource usage increases. The static strategies (centralized and static K-means) are not included in the figure, since they do not involve any deployment overhead (the same anchor points are permanently deployed throughout the simulation). As expected, the random algorithm introduces the highest overhead, by requiring almost a complete re-deployment of the whole set of anchor points at each time slot. The K-means strategy introduces the second highest deployment overhead, by requiring a re-deployment of about 60% of the anchor points in different locations. The K-means greedy average and the modularity greedy average are slightly better than the K-means strategy. These clustering-oriented strategies re-deploy half of the anchor points in different locations at every time slot. Next, the greedy average and greedy percentile strategies introduce very low deployment overheads, less than 15% of re-deployed anchor points with greedy average and about 5% with greedy percentile. Finally, the overhead-aware greedy average achieves an even greater reduction, lower than 2% of the deployment overhead.

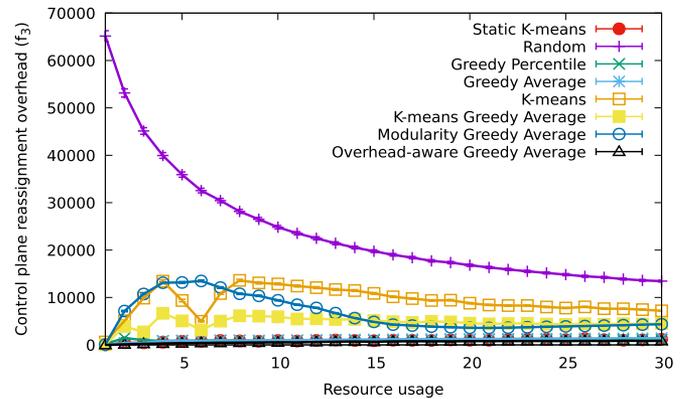

Fig. 9. Control-plane reassignment overhead introduced in the network ($f_3$, overhead in terms of the cost for relocating vehicle communications) as resource usage ($N_{\text{anchor\_points}}$, number of anchor points deployed) increases.

Fig. 9 shows the control-plane reassignment overhead $f_3$ in (5) as the resource usage increases. The centralized algorithm is not shown in the figure because it does not involve any reassignment (i.e., all the UEs are assigned to the single core anchor point). We can observe that the random algorithm leads to the highest control-plane reassignment overhead. The curve for this algorithm decreases as the resource usage increases. The reason for this behavior is also related to the deployment overhead: since most deployed anchor points change at each time slot, all the users must be reassigned to a different anchor point. Besides, when anchor points are randomly selected, low resource usages imply that the distance between the previously deployed anchor points and the newly deployed ones will be large. As the resource usage increases, the distance between the anchor points that were previously deployed and the newly deployed ones decreases, and so does the control-plane reassignment overhead. Regarding the alternative algorithms, K-means introduces the higher values, except for a resource usage of around 6 anchor points, for which the overhead decreases notably, reflecting the higher instability of this algorithm. The K-means greedy average and the modularity greedy average algorithms are better than K-means. However, for low resource usages (less than 15 anchor points deployed), the modularity greedy average is comparable to K-means. High fluctuations can also be appreciated in the curve of the K-means greedy average algorithm for resource usages of less than 10 anchor points. Clearly, the overhead-aware greedy average provides the lowest control plane reassignment overhead, for a very stable behavior throughout the simulation. The greedy average, greedy percentile and static K-means provide are slightly worse and exhibit non-negligible fluctuations, especially at low resource usages. Fig. 10 shows a detailed view of the control-plane reassignment overhead, for the static K-means, greedy percentile, greedy average and overhead-aware greedy average algorithms.

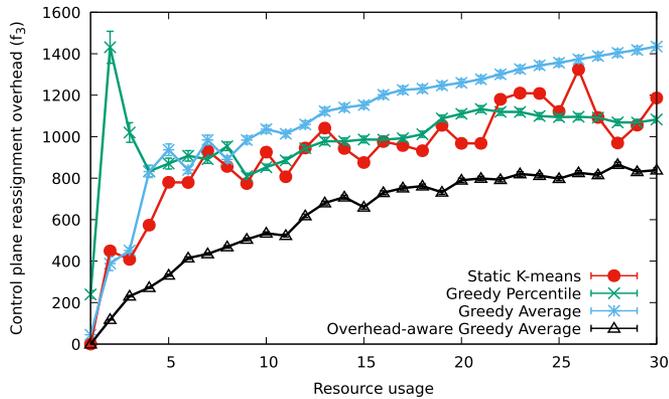

Fig. 10. Control-plane reassignment overhead introduced in the network ($f_3$, overhead in terms of the cost for relocating vehicle communications) as resource usage ($N_{\text{anchor\_points}}$, number of anchor points deployed) increases (static K-means, greedy percentile, greedy average and overhead-aware greedy average algorithms).

In practical scenarios, a proper resource usage can be derived from the latency results in Fig. 6, depending on the target latency requirements. As an example, if we consider stringent requirements for vehicular communications of 5 ms [37], by assuming link latencies of 1 ms, the centralized deployment would not be feasible, since the latency would exceed 20 ms. Other baseline strategies, the random and static K-means baselines, would respectively require the deployment of 25 and 17 anchor points. The greedy average, greedy percentile and overhead-aware greedy average algorithms can achieve the goal by deploying 10 anchor points. Interestingly, the proposed overhead-aware greedy average algorithm is able to attain similar levels of latency performance while also reducing the deployment and control-plane reassignment overheads. A possible limitation of this algorithm is execution time, which is almost 800 ms for the studied scenario, while the greedy average takes about 100 ms. Moreover, $\alpha_i$ parameters should also be selected to attain the desired trade-off level between the different objective functions. In our tests, values $\alpha_1 = 0.5$, $\alpha_2 = 0.25$ and $\alpha_3 = 0.25$, achieved a balancing trade-off between the latency and the overheads.

## VII. CONCLUSION

Cellular networks are gaining attention in automotive scenarios such as connected vehicles. Low latency is one of their major requirements, especially for safety-related applications. The MEC paradigm is a key enabling technology for satisfying this requirement in 5G and 6G cellular networks.

In this article we have addressed the problem of dynamically deploying anchor points in MEC hosts and assigning vehicular UEs to them for jointly reducing the latency perceived by the vehicles and the overhead introduced by network reconfigurations, such as anchor point redeployments and vehicle UEs' reassignments to anchor points. We have formally defined the problem as a multi-objective ILP optimization model. We have proposed a novel heuristic algorithm for solving it and compared its performance with baseline strategies and previous latency-minimization strategies. We have analyzed the trade-off between resource usage, the latency perceived by the UEs, the overhead introduced in the network and the running time of the algorithms.

Our results show that our anchor point deployment and terminal assignment algorithm achieves satisfactory trade-offs between the goals of the optimization model. It attains latency levels that are comparable to those of the competing latency minimization algorithms while consistently reducing the network overhead. Overall, our proposal can help to integrate latency-sensitive vehicular applications into cellular networks with few network reconfigurations. As future work, we will study clustering-based strategies to improve the scalability of the overhead-aware greedy average algorithm.

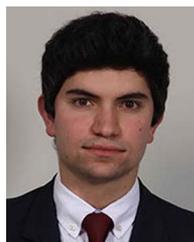

**Pablo Fondo-Ferreiro** received the bachelor's and master's degrees in telecommunication engineering from the University of Vigo, Vigo, Spain, in 2016 and 2018, respectively, and the Ph.D. Degree in information and communication technologies from the University of Vigo, funded by a la Caixa Foundation Fellowship. He is currently a Postdoctoral Researcher with the University of Vigo funded by a Xunta de Galicia Fellowship. In 2022, he joined the Instituto de Telecomunicaçoes, Aveiro, Portugal, for a 2-year postdoctoral research stay. He has participated in ten competitive research projects and co-authored more than ten papers in peer-reviewed journals and international conference proceedings. His research interests include SDN, mobile networks, MEC, and artificial intelligence. He was the recipient of the award for the best academic record.

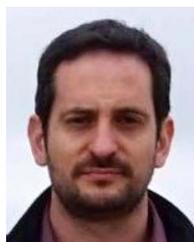

**Felipe Gil-Castiñeira** received the M.Sc. and Ph.D. degrees in telecommunication engineering from the University of Vigo, Vigo, Spain, in 2002 and 2007, respectively. He is currently an Associate Professor with the Department of Telematics Engineering, University of Vigo. Between 2014 and 2016, he was the Head of advanced telecommunications with iNetS Area, Galician Research and Development Center. He has authored or coauthored more than 60 papers in international journals and conference proceedings. He has led several national and international R&D projects. He holds two patents in mobile communications. His research interests include wireless communication and core network technologies, multimedia communications, embedded systems, ubiquitous computing, and the Internet of things. He is the Co-founder of an University spin-off, Ancora.

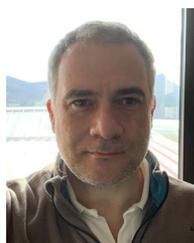

**Francisco Javier González-Castaño** is currently a Catedrático de Universidad (Full Professor) with the Telematics Engineering Department, University of Vigo, Vigo, Spain, where he leads the Information Technology Group. He has authored more than 100 papers in international journals in the fields of telecommunications and computer science, and has participated in several relevant national and international projects. He holds 3 U.S. patents.

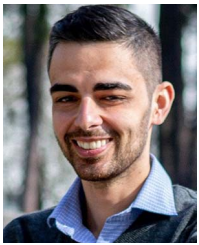

**David Candal-Ventureira** received the bachelor's and master's degrees in telecommunication engineering from the University of Vigo, Vigo, Spain, in 2016 and 2018, respectively. He is currently working toward the Ph.D. Degree in information and communication technologies. Since 2018, he has been a Researcher with the Information Technologies Group, University of Vigo. His research interests include mobile and wireless networks, and artificial intelligence.

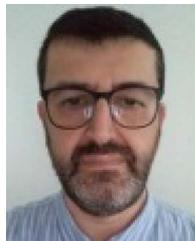

**Antonio J. Morgado** received the graduation degree in electronic and telecommunications engineering from the University of Aveiro, Aveiro, Portugal, in 1997. He is currently working toward the Doctoral degree with the University of South Wales, Newport, U.K. Since then, he was with Instituto de Telecomunicações (IT-Aveiro), Aveiro, where he was involved in several Portuguese and European (FP5, FP6) research projects. He was a Teaching Assistant with the University of Aveiro, Portugal from 2003 to 2011, and with the Polytechnic Institute of Viseu, Viseu, Portugal, from 2011 to 2012. He returned to IT-Aveiro in 2012, to perform research on TV white spaces. From 2014 to 2016, he participated in European FP7- ADEL project dealing with licensed shared access. His research interests include spectrum regulation, machine learning, 5G/6G standardization, radio resource management, D2D, and mmWave communications.

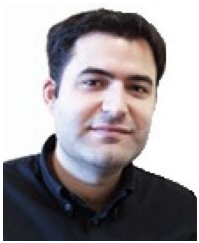

**Jonathan Rodriguez** (Senior Member, IEEE) received the Ph.D. degree from the University of Surrey, Guildford, U.K., in 2004. In 2005, he became a Researcher with the Instituto de Telecomunicações, Aveiro, Portugal, and Senior Researcher in 2008. He has was the Project Coordinator for major international research projects (Eureka LOOP, FP7 C2POWER, and H2020-MSCA-SECRET), whilst acting as the Technical Manager for FP7 COGEU and FP7 SALUS. He is currently the Project Director for the NATO SPS-PHYSEC Project targeting physical layer security. He is the author of more than 600 scientific works, that include 11 book editorials, and currently an Associate Editor for IEEE ACCESS and *IET Communications journal*. Since 2017, he has been a Professor of mobile communications with the University of South Wales, Newport, U.K. Since 2013, he has been a Chartered Engineer. In 2015, he was a Fellow of IET.

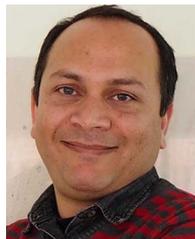

**Shahid Mumtaz** (Senior Member, IEEE) is currently with the Department of Applied Informatics, Silesian University of Technology, Gliwice, Poland, and Departement of Engineering, Nottingham Trent University, Nottingham, U.K. His work has resulted in technology transfer to companies and patented technology. His expertise lies in 5G/6G wireless technologies using AI/ML and digital twin (VR/XR) tools and innovation paths in industry and academia. Moreover, he was a Senior 5G consultant with Huawei and InterDigital, where he contributed to RAN1 /RAN2 and looked after the university-industrial collaborative projects. He is an IET Fellow, an IEEE ComSoc, and ACM Distinguished Lecturer. He was the Founder and Editor-in-Chief of *IET's Journal of Quantum Communication*, Editor-in-Chief of the *Alex-andria Engineering Journal* (Elsevier), ViceChair, Europe/Africa Region - IEEE ComSoc Green Communications Computing Society, and Vice-Chair for IEEE Standard P1932.1, Standard for Licensed/Unlicensed Spectrum Interoperability in wireless mobile networks. He was the recipient of the IEEE ComSoc Young Researcher Award.